\documentclass[showpacs,aps,prb,twocolumn]{revtex4-1}
\usepackage{amsmath}
\usepackage{amsfonts}
\usepackage{amssymb}
\usepackage{amsthm}
\usepackage{graphicx}
\usepackage{bbm}
\usepackage{bbold}
\usepackage[T1]{fontenc}
\usepackage[mathscr]{euscript}
\newcommand{\sinc}{{\rm sinc}}

\begin{document}


\title{Photoinduced pseudospin effects in silicene beyond the off resonant condition}
 \author{Alexander L\'{o}pez$^{1,2}$}
 \email[To whom correspondence should be addressed. Electronic
 address: ]{alexander.lopez@physik.uni-regensburg.de}
\author{Andreas Scholz$^{2}$}
\author{Benjamin Santos$^{3}$}
\author{John Schliemann$^{2}$}
\affiliation{1. School of Physics Yachay Tech, Yachay City of Knowledge 100119-Urcuqui, Ecuador\\
2. Institute for Theoretical Physics, University of Regensburg, D-93040
Regensburg, Germany\\
 3.INRS-EMT, Universit\'e du Qu\'ebec, 1650 Lionel-Boulet, Varennes, Qu\'ebec J3X 1S2, Canada
 }
\date{\today}

\begin{abstract}
We study the photoinduced manipulation of charge carriers in monolayer silicene subject to intense electromagnetic  terahertz radiation. 
Considering the Dirac cone approximation and going beyond the off resonant condition for large frequencies of the radiation field, 
where only virtual photon processes are allowed, we present the exact zero-momentum 
pseudospin polarization and numerical results for the quasienergy band structure and time-averaged density of states. We find that  
resonant processes, due to real photon emission and absorbtion processes, induce a band inversion that qualitatively modifies the quasienergy spectrum. These band structure changes 
manifest themselves as an inversion of the averaged pseudospin polarization.  Through the analysis of the time-averaged density of 
states we find that effective photoinduced gap manipulation 
can only be achieved in the intermediate and strong matter-radiation coupling regime where the off resonant approximation breaks down.
\end{abstract}
\maketitle
\section{Introduction}
The dynamical control of the electronic properties of Dirac fermions in the solid state environment by means of time periodic fields is currently an intense research topic.
\cite{fti,foa,foa2} Among the two-dimensional materials supporting these Dirac fermions we have as prominent examples graphene\cite{novoselov1,geim,guinearmp} and silicene\cite{silicene1,silicene2}. 
In this work we focus our attention on silicene which consists of a two-dimensional honeycomb lattice structure made of silicon atoms analogous to that of graphene. 
From the experimental point of view some recent works have reported the synthesis of silicene\cite{silicene3,silicene4,silicene5}. As well as in the case of graphene, the silicene 
honeycomb lattice consists of two triangular sublattices. However, silicene has a corrugated or buckled lattice structure 
that makes the silicon atoms in one sublattice to be perpendicularly displaced with respect to those atoms lying on the other 
sublattice. In the low energy Hamiltonian description of silicene, this sublattice degree of freedon is formally associated to a 
quantity called pseudospin which resembles the real spin. Moreover, in momentum space 
there are two degenerate energy extrema called Dirac points, denoted by momenta $\pm K$, that are related by time reversal 
symmetry and they lie at opposite corners of the hexagonal Brillouin zone. 
To this energy extrema one can associate a valley degree of freedom which in turn can also be described as a pseudospin\cite{guinearmp}. 
This degree of freedom  has been shown 
to be suitable for the potential realization of {\it valleytronics}, i.e., the analogous to spintronics based on the real spin
(for a review on spintronics see \cite{zutic}).\\
\noindent In the case of the sublattice pseudospin there have also been proposals to realize the so called 
{\it pseudospintronics}, where physical operations such as pseudospin magnetism in bilayer graphene\cite{min} can in principle be performed by means of this physical quantity. 
This in turn steems from the chiral nature of the Hamiltonian eigenstates for which the pseudospin is locked to the charge carrier's momentum.
This chirality has profound consequences that include an unusual sequencing of plateaus in measurements of the quantum
Hall effect\cite{uno}. In addition, in the conduction band of valley $K$ pseudospin is parallel to the momentum while in the valence band, 
pseudospin is antiparallel to the charge carriers momenta. Therefore, another physical manifestation of this pseudospin degree of freedom in graphene is that chiral states
can be perfectly transmitted through a potential barrier which constitutes a realization of the
Klein paradox in condensed matter\cite{novoselov-klein}. In silicene, another pseudospin effect has been predicted to appear
when a perpendicular electric field $E_z$ is applied since, in this case, the atoms belonging to 
each sublattice would respond differently to $E_z$, giving rise to an staggered potential\cite{ezawa1}. Due to this peculiar pseudospin 
response to applied electric fields,  the electronic properties of silicene are predicted to considerably differ from those 
of graphene, despite their formal similarities. In particular, one can induce a pseudospin polarization in the silicene sample
by means of a perpendicular static electric field. Since the pseudospin degree of freedom mus be included in the total angular
momentum operator\cite{regan}, this pseudospin polarization can be interpreted as a differential population
of the charge carriers on each sublattice as a response of the charge carriers to the angular momentum content of the 
circularly polarized radiation field. Moreover, the linear spectrum of the low energy Hamiltonian (near the Dirac points) for 
both graphene and silicene leads to a Fermi velocity that is independent of momentum. In fact, within the Dirac cone approximation,
the velocity operator is proportional to the pseudospin operator describing the sublattice degree of freedom\cite{guinearmp}.  
 In presence of a radiation field, the pseudospin gets coupled (via the minimal prescription) 
to the electromagnetic field, and thus, dynamical modulation has been predicted to appear both in graphene\cite{foa} and silicene\cite{ezawa2} either 
at zero or finite momentum.\cite{andreas}

\noindent Another interesting feature of silicene is that its intrinsic spin-orbit coupling is much larger than that of pristine 
graphene. Therefore, an interesting interplay among intrinsic spin-orbit coupling and electric field effects was predicted to appear because the bandgap can be electrically controlled. Moreover, the addition of 
an exchange potential term (which physically could represent the proximity effect due to coupling to ferromagnetic leads) allows for topological quantum phase transitions in the static regime.\cite{ezawa1} Furthermore, in presence of circularly polarized electromagnetic radiation it has 
been recently proposed the realization of the so called single Dirac cone phase\cite{ezawa2}. At this topological phase, it is found that well defined spin polarized states are supported at every Dirac 
point. Within this configuration, different spin components propagate in opposite directions giving rise to a pure spin current at finite momentum.\cite{ezawa2} Yet, these photoinduced topological 
phase changes\cite{kane,z2,rmp,topological1,topological2} reported by Ezawa\cite{ezawa2} were derived under the off resonant assumption, i.e., dynamical processes such that
the frequency (coupling strength) of the radiation field is much larger (smaller) than any other energy scale in the problem. Under these assumptions it is possible to derive an effective 
time-independent Floquet Hamiltonian\cite{milena,chu} with a tiny photoinduced bandgap correction that stems from 
virtual photons that dress the static energy eingenstates. Since the sign of the bandgap  
term  (i.e., the effective bandgap) determines important topological properties of the material, it is vital both for potential practical implementations, for instance in techonological 
realizations of silicene-based devices, as well as from a fundamental point of view, to effectively achieve manipulation of this quantity.\\ 

\noindent In this 
work we show that in order to detect relevant photoinduced effects in the band structure of silicene under strong circularly polarized electromagnetic radiation in the terahertz 
(frequency) domain one needs to go beyond the aforementioned off resonant approximation. At intermediate coupling regime we reproduce the single valley Dirac phase reported by Ezawa\cite{ezawa2} and we show that 
effective dynamical gap closing occurs at or above the intermediate coupling regime of the Dirac fermions to the radiation field. By exact evaluation of the zero-momentum pseudospin polarization we find that
pseudospin inversion can only be dynamically achieved at intermediate or strong coupling of the charge carriers to the radiation field and thus, the off resonant
modifications induced in the band structure turn out to be a rather small effect. This is verified by a numerical evaluation of the 
time-averaged density of states.
\\

\noindent The paper is organized as follows. In section II we present the model and obtain the quasienergy spectrum along with the exact zero-momentum dynamical polarization. In section III we present our results for the finite momentum quasienergy spectrum as well as the Density of States (DoS). In section IV we discuss 
the main results and give some concluding remarks.
\section{model}
We adopt the Dirac cone approximation to describe the dynamics of non interacting charge carriers in silicene subject to a perpendicular, uniform and constant electric field 
${\bf E}=E_z\hat{z}$. This is given by the $8\times8$ Hamiltonian\cite{ezawa1}  ($\hbar=e=1$, with $e$  being the 
electron's charge) 
  \begin{eqnarray}\label{e1}
\mathcal{H}^\eta&=&v_F( k_x\sigma_x+\eta k_y\sigma_y)+\sigma_z(\eta s_z\lambda_{so}-\ell E_z)\\\nonumber 
&&+\eta\sigma_zh_{11}+h_{22}
\end{eqnarray}
where $v_F=\frac{\sqrt{3}at_b}{2}\approx 8.1\times10^5\,\textrm{m/s}$ is the Fermi velocity for charge 
carriers in silicene, with  $a = 3.86$\, \AA\,  the lattice constant and $t_b=1.6\, \textrm{eV}$ the hopping parameter within a tight-binding formulation, whereas 
$\ell=0.23\,$\AA\, measures half the separation among the two sublattice planes. 
In addition, $\eta=\pm 1$ describes the Dirac point, $\sigma_i$ and $s_i$ ($i=x,y,z$) are Pauli 
matrices describing pseudo and real spin degrees of freedom, respectively, whereas the time reversal symmetry of the two Dirac points
can be encoded in the momentum as $\vec{k}=(k_x,\eta k_y)$, i.e., it is the momentum 
measured from the corresponding Dirac point $\eta=\pm1$. Following reference, we are using a cooordinate system with the $x$ axis being 
perpendicular to the two inequivalent silicon atoms in the unit cell. The parameter $\lambda_{so}=3.9\, \textrm{meV}$ represents the strength of the 
intrinsic spin-orbit contribution. Moreover, the two  contributions given by the terms
\begin{eqnarray}
h_{11}&=& a\lambda_{R2}(k_ys_x-k_xs_y),\\ 
h_{22}&=& \lambda_{R1}(\eta\sigma_x s_y-\sigma_ys_x)/2, 
\end{eqnarray}
describe the spin-orbit coupling associated to the next nearest 
neighbor hopping and nearest neighbors tight binding formulation, respectively.\\

\noindent The term $h_{11}$ has its origin in the buckled 
structure of silicene whereas $h_{22}$ is induced by the application of an external static electric field $E_z$. 
Using first principle calculations, the authors of reference \cite{sili-rashba} found that $\lambda_{R1}=0.2\, \textrm{meV}$ for a typical
electric field $E_z= (50V)/300nm$ whereas 
$h_{22}$ is of order $10\mu\textrm{eV}$ for a critical electric field $E_c=\lambda_{so}/\ell=17\textrm\,{meV}$. In this manner,
$h_{22}$ is much smaller than the other energy scales in the problem. Therefore, these two non conserving contributions will be neglected in the following, although in the appendix we show that the largest 
contribution $h_{11}$ can be easily incorporated in the solution to the dynamical evolution presented below. Yet, we have verified that our results do not qualitatively change by the introduction 
of these two small corrections. 

Within the approximation $h_{22}=0$, let us now consider the  pseudospin  dynamics under an {\it intense} radiation field represented by the time-dependent vector potential
\begin{equation}
{\bf A}(t)=A\left(\cos\Omega t,\sin\Omega t\right),
\end{equation}
with $A=\mathcal{E}/\Omega$ and $\Omega$ its amplitude and frequency, respectively. It describes a monochromatic electromagnetic wave incident perpendicular to the sample. This vector potential 
can in turn be derived from the corresponding electric field by means of ${\bf E}(t)=-\partial_t {\bf A}(t)$, where $\mathcal{E}$ is the amplitude of the time-dependent electric field. 

Using the standard minimal coupling prescription given as $\vec{k}\rightarrow \vec{k}+\vec{A}$, we get the dynamical generator
\begin{eqnarray}\label{e11t}
\nonumber\mathcal{H}^\eta(\vec{k},t)&=&v_F(\sigma_xk_x+\eta\sigma_y k_y)+\sigma_z(\eta s_z\lambda_{so}-\ell E_z)\\ 
&&v_FA[\sigma_x\cos\Omega t+ \eta\sigma_y\sin\Omega t].
\end{eqnarray}
In the following we will explore the emerging photoinduced dynamical features at different momentum scenarios. For this purpose, we explore the low, intermediate and strong coupling regimes
of the charge carriers in silicene under the radiation field. 
\subsection{Physics at $\textrm{k}=0$}
At zero momentum the extrinsic spin-orbit term $h_{11}$ vanishes and the z-component of spin $s_z=\pm1$ is a good quantum number. Therefore, the following analysis is independent of taking into account the 
aforementioned spin-orbit contribution. Setting for notational convenience $\alpha=v_FA$ and $V_z=\ell E_z$, the physics at zero momentum $\vec{k}=0$ is described by the dynamical generator
\begin{eqnarray}\label{e1t}
\mathcal{H}^\eta(0,t)&=&\nonumber(\eta s\lambda_{so}-V_z)\sigma_z+\alpha[e^{i\eta\Omega t}\sigma_-+ e^{-i\eta \Omega t}\sigma_+].\\
\end{eqnarray}
From this equation we note that the sublattice degree of freedom must be included in the total angular momentum of the system 
in order to account for conservation of this quantity as a consequence of the rotational invariance of the system that is preserved
in absence of Rashba spin-orbit terms. This was another motivation for studying the zero-momentum pseudospin modifications induced
by the radiation field. Now if we apply the unitary transformation
\begin{equation}
\mathcal{P}^\eta(t)=e^{-i\eta(\mathbb{1}+\sigma_z)\Omega t/2} 
\end{equation}
we get the effective time-independent Floquet Hamiltonian $\mathcal{H}_F(k=0)=(\mathcal{P}^\eta)^\dagger(t)\mathcal{H}^\eta(0,t)\mathcal{P}^\eta(t)-i(\mathcal{P}^\dagger)^\eta(t)\dot{\mathcal{P}}^\eta(t)$
\begin{eqnarray}\label{flo0}
\mathcal{H}_F(k=0)&=&\nonumber -\frac{\eta\Omega}{2}\mathbb{1}+\Bigg[\eta\Bigg( s_z\lambda_{so}-\frac{\Omega}{2}\Bigg)-V_z\Bigg]\sigma_z+\alpha\sigma_x.\\
\end{eqnarray}
\begin{figure*}[ht]
\begin{tabular}{cc}
\includegraphics[height=6cm]{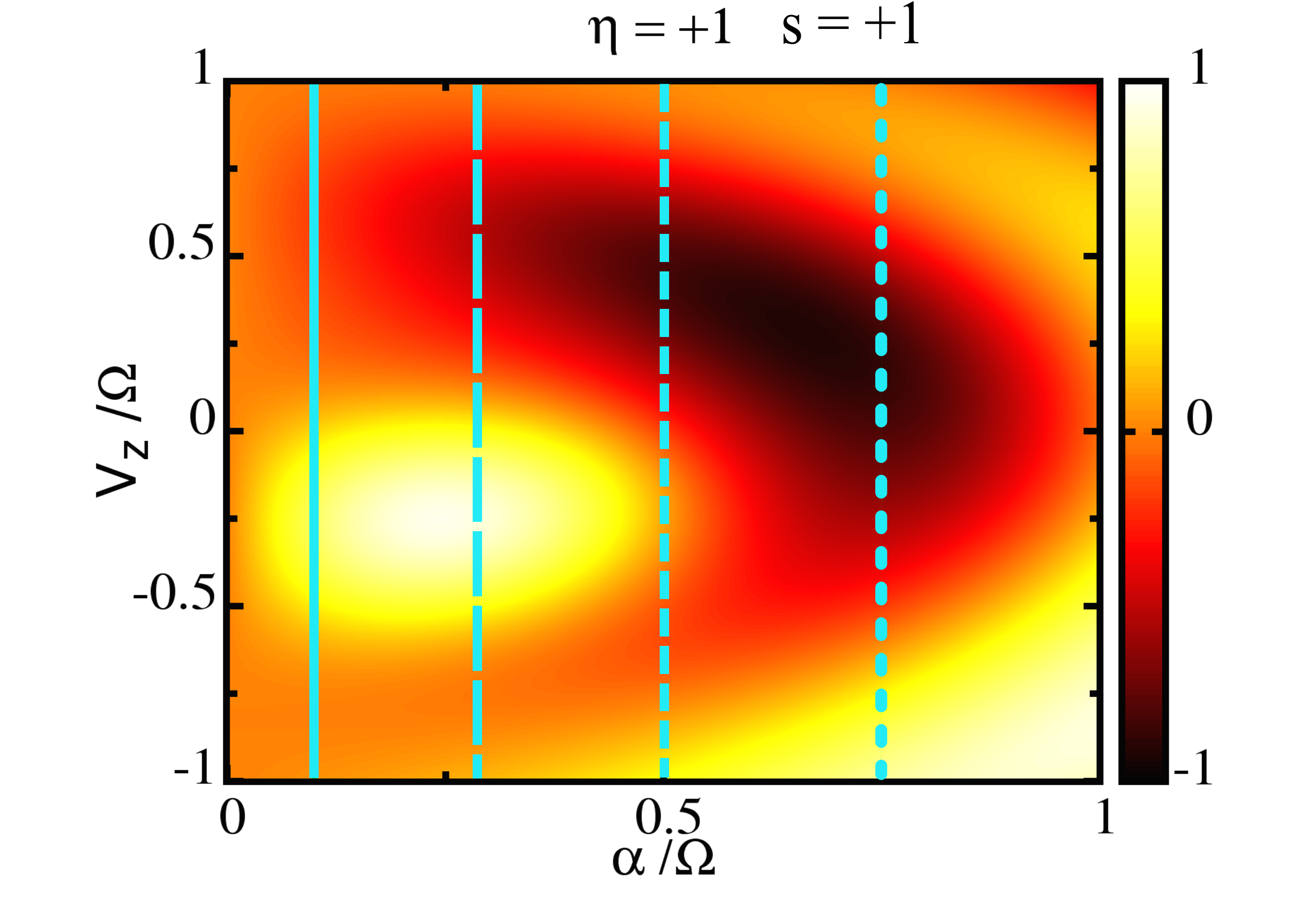}&
\includegraphics[height=6cm]{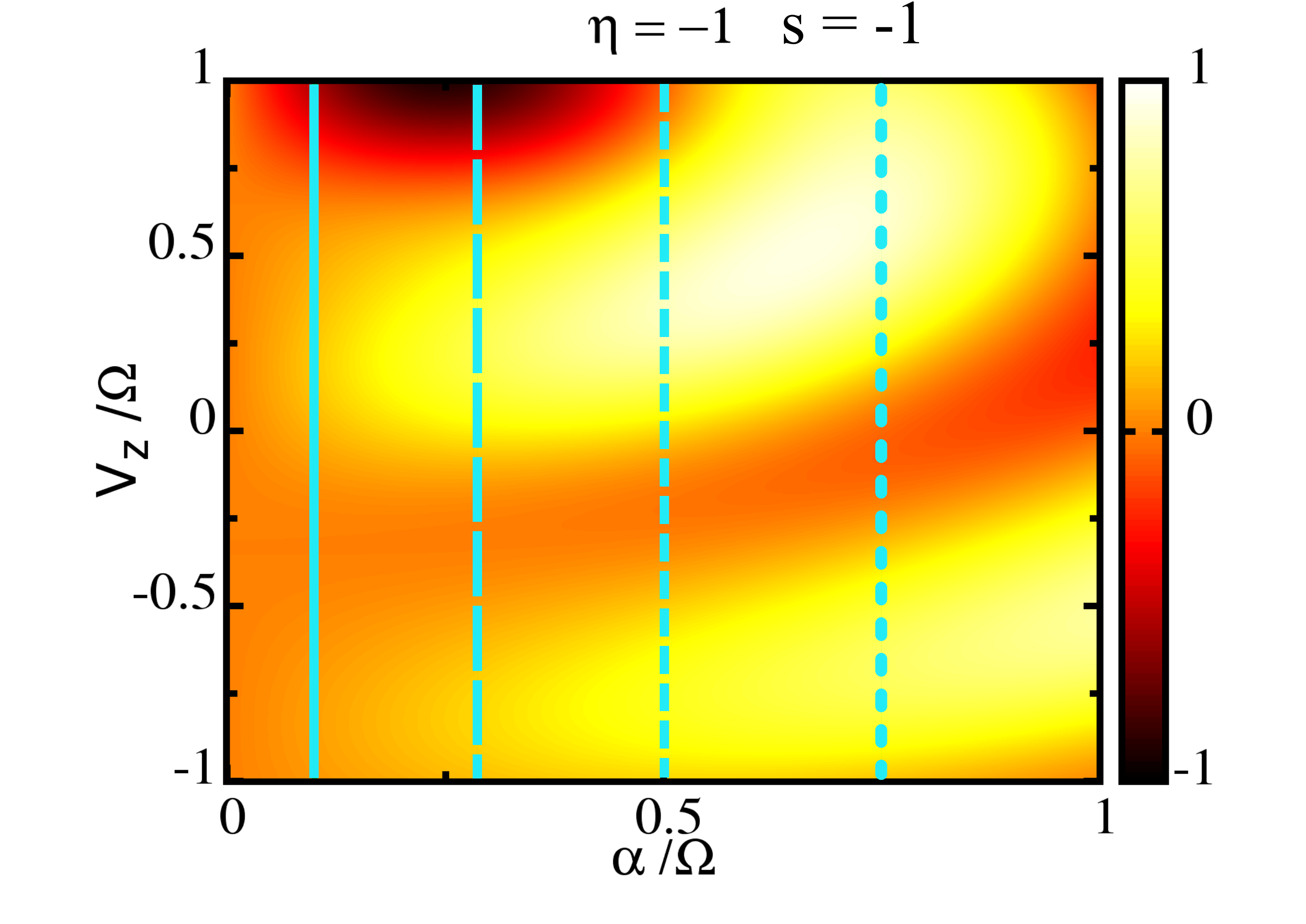}\\
\includegraphics[height=6cm]{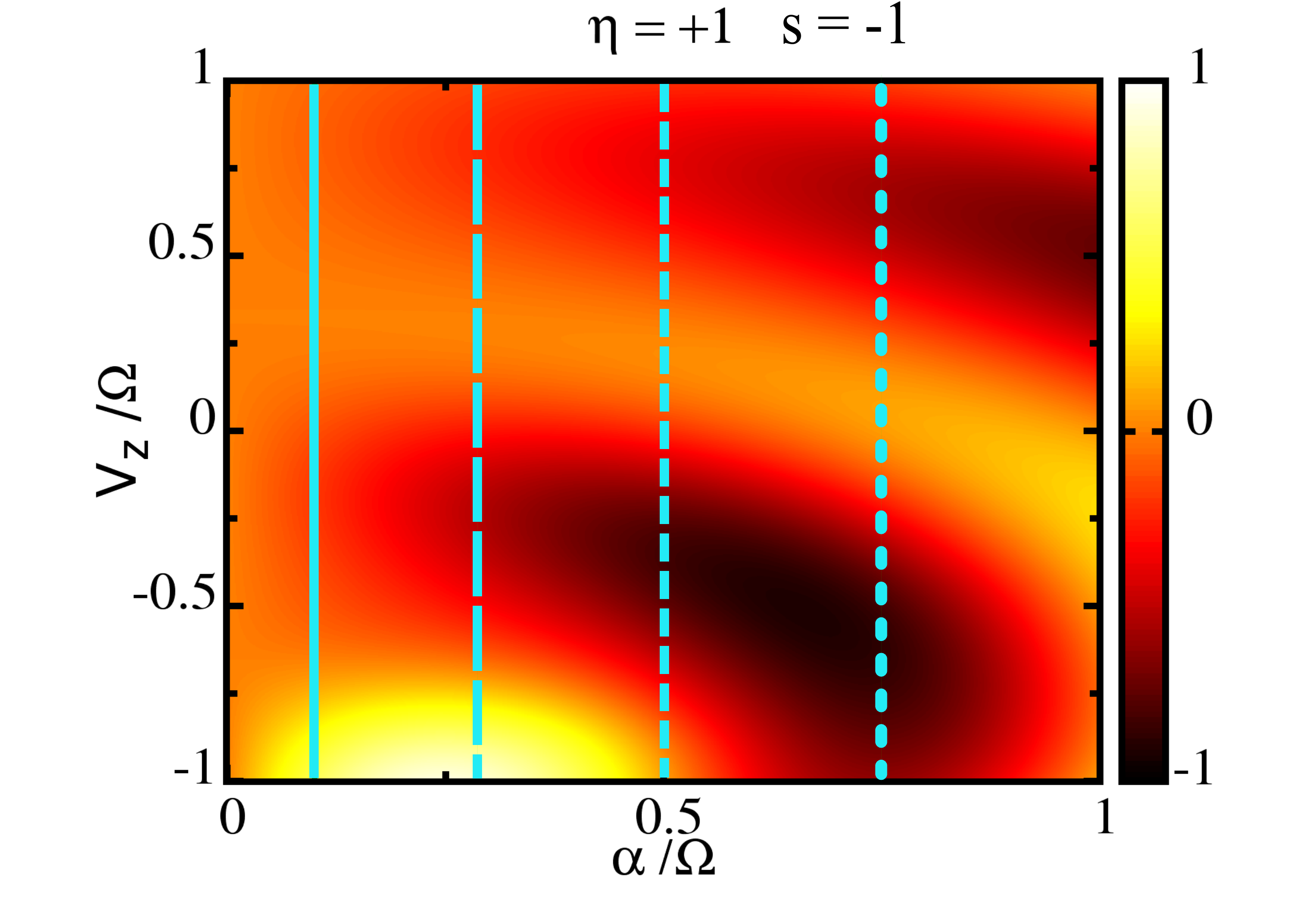}&
\includegraphics[height=6cm]{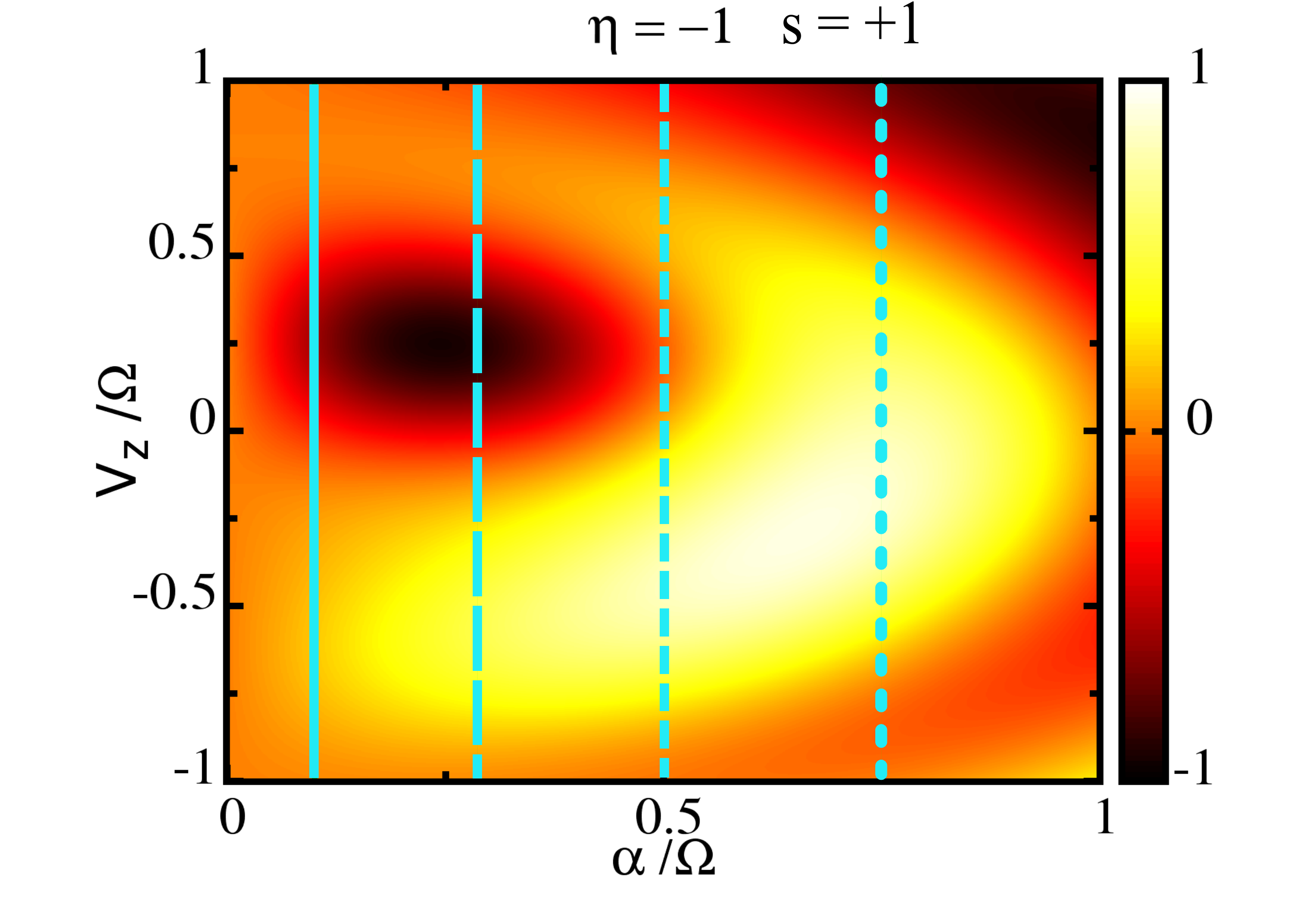}
\end{tabular} 
 \caption{\label{fig:figure1}(Color online) Zero momentum pseudospin mean polarization $\langle\sigma_z\rangle$ as given in equation $(16)$, for $\theta=\pi/2$ and
 $\phi=\pi/4$, for different combinations of the product $s\eta$. The vertical light blue lines correspond to $\alpha=0.1\Omega$ (continuos), $\alpha=0.25\Omega$ (large dashed)
 $\alpha=0.5\Omega$ (short dashed) and $\alpha=0.75\Omega$ (large dots), respectively. In this and next figures, we have set a frequency in the far infrared domain 
 $\Omega=3\textrm{THz}$. See discussion in the main text.}
\end{figure*}
Thus, the static Floquet Hamiltonian (\ref{flo0}) shows that the radiation field couples in a non diagonal form to the pseudospin degree of freedom through the last term and therefore, can induce pseudospin
dynamical modulation, even at zero momentum. The Hamiltonian (\ref{flo0}) resembles that of the Rabi problem for a real spin
in an external oscillating magnetic field. Therefore, the radiation field could be used to coherently control the pseudospin degree of freedom in analogy 
to the coherent manipulation of the real spin by means of electric fields in GaAs semiconducting quantum dots\cite{coherent}. 
To explicitly show this, we find the zero momentum quasienergy spectrum wich is given as
\begin{equation}
\varepsilon^\eta_{s\sigma}(k=0)=-\frac{\eta\Omega}{2}+\sigma\sqrt{\alpha^2+(\Delta^\eta_s)^2},
\end{equation}
where $s,\sigma=\pm 1$ represent the real and pseudospin degrees of freedom, respectively. In addition, we have defined the effective gap 
\begin{equation}
\Delta^\eta_s=\eta\Big(s\lambda_{so}-\frac{\Omega}{2}\Big)-V_z.
\end{equation}
We can also introduce the Rabi frequency, defined as $\Gamma=\sqrt{\alpha^2+(\Delta^\eta_s)^2}$, that would dictate the coherent 
oscillations between the two static pseudospin eigenstates of $\sigma_z$.
On the other hand, the zero-momentum exact Floquet eigenstates are 
\begin{equation}\label{basis00}
 |\psi^\eta_{s\sigma}(t)\rangle=\frac{e^{-i\varepsilon^\eta_{s\sigma}t}}{\sqrt{2\Gamma}}\left(
\begin{array}{c}
e^{-i\eta\Omega t}\sqrt{\Gamma+\sigma\Delta^\eta_ s}\\
\sigma \sqrt{\Gamma-\sigma\Delta^\eta_s}
\end{array}
\right),
\end{equation} 
In order to analyze the dynamical manipulation of the pseudospin degree of freedom, let us now assume that the system is 
initially prepared in the arbitrary state 
\begin{equation}\label{phi0}
 |\Phi(0)\rangle=\left(
\begin{array}{c}
\cos\frac{\theta}{2}e^{i\phi/2}\\
\sin\frac{\theta}{2}e^{-i\phi/2}
\end{array}
\right),
\end{equation}
\begin{figure*}[ht]
\begin{tabular}{cc}
\includegraphics[height=6.25cm]{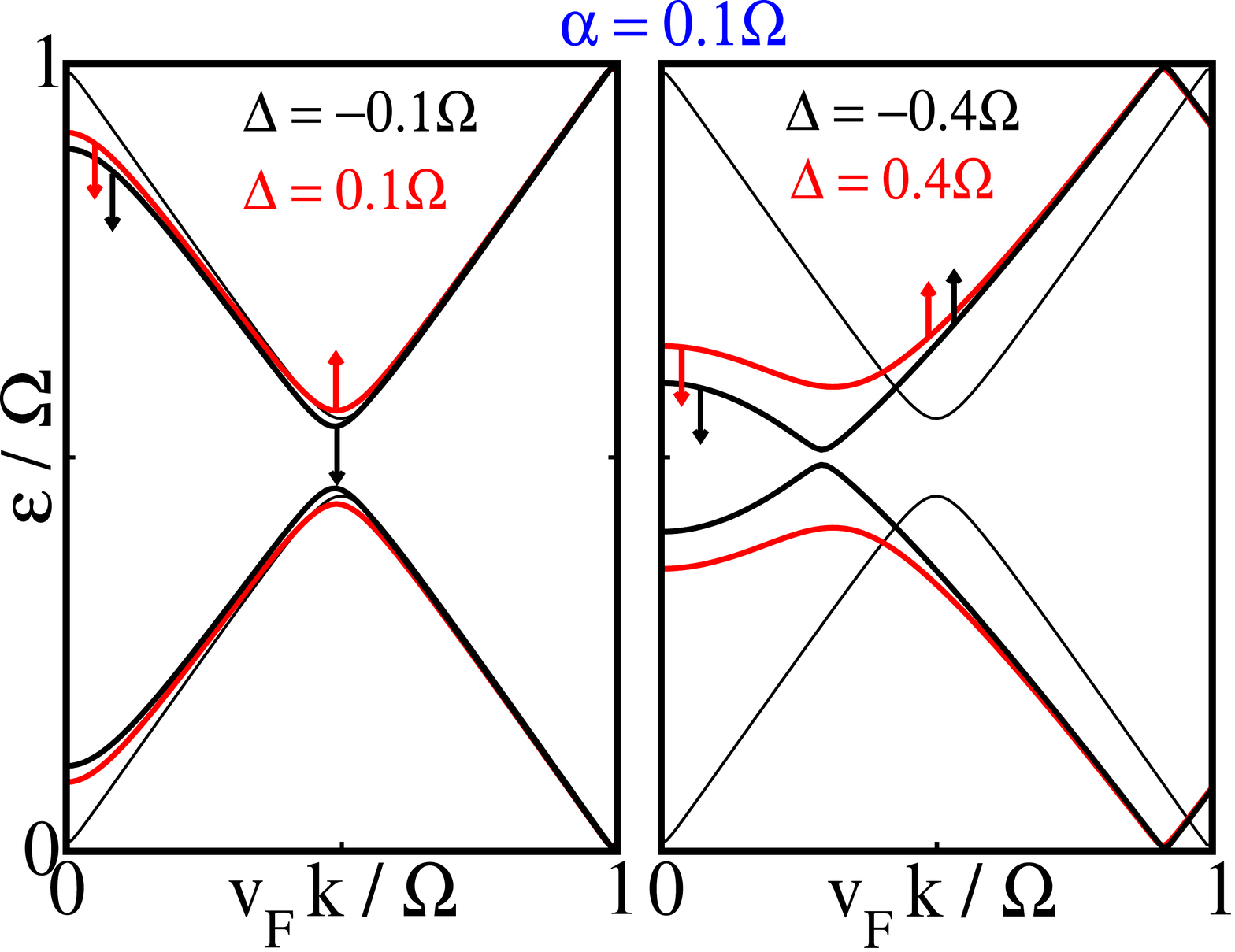}&
\includegraphics[height=6.25cm]{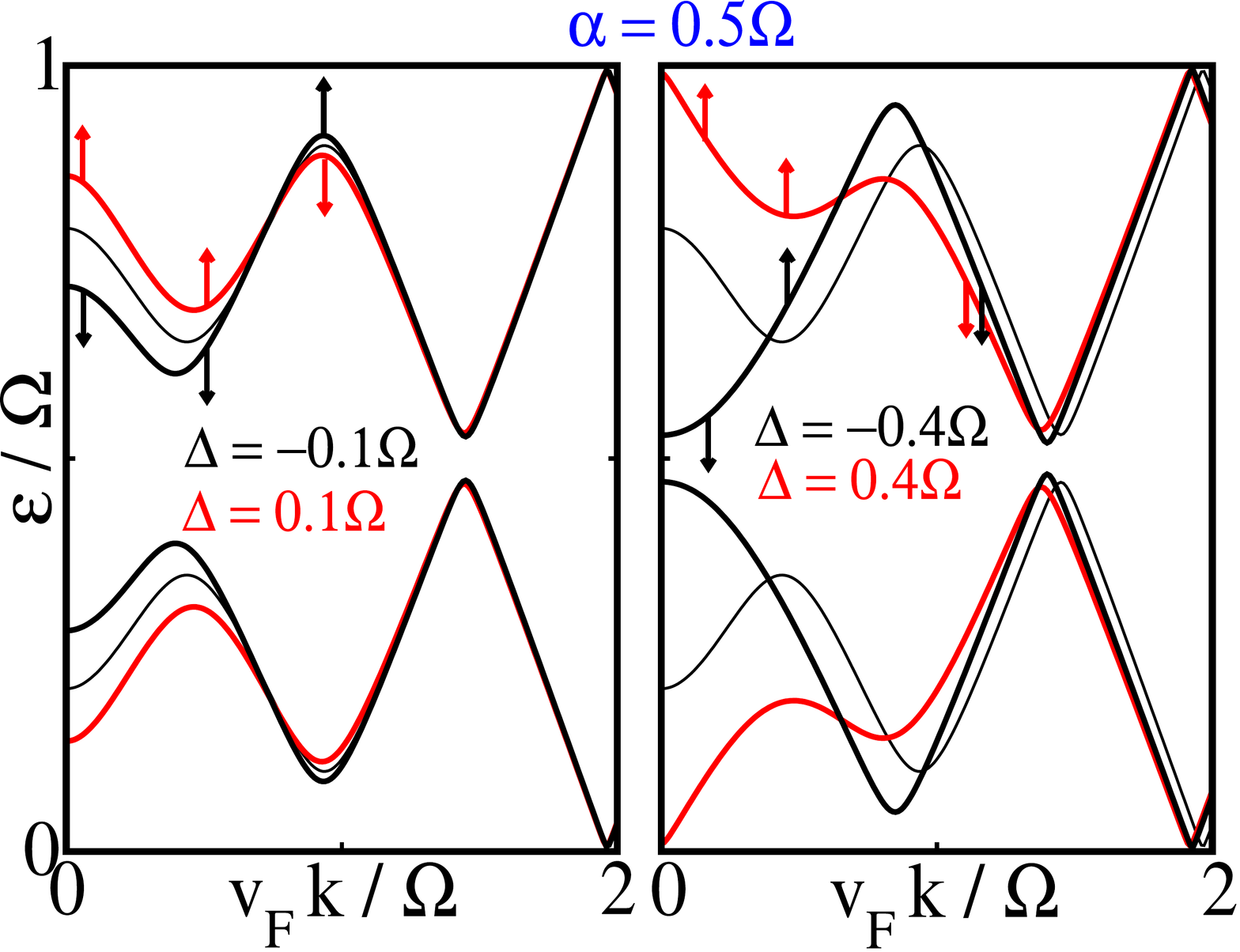}
\end{tabular}
\caption{\label{fig:figure2}(Color online) Momentum dependent quasienergy spectra, within the first Brillouin zone $0<\varepsilon<\Omega$,
at low (intermediate) $\alpha=0.1\Omega$ ($\alpha=0.5\Omega$) light-matter coupling values. We consider positive (red, thick curves) and negative 
(black, thick curves) for both small $\Delta=0.1\Omega$ and intermediate $\Delta=0.4\Omega$ absolute values of the static bandgap. The black thin lines correspond to 
$\Delta=0$.  As a guide to the eye we have used red (black) arrows that point ``away'' from the $\Delta=0$ quasienergy 
spectrum for positive (negative) values of $\Delta$.}
\end{figure*} 
with $0\le\theta\le\pi$ and $0<\phi\le2\pi$ being spherical coordinates over the Bloch sphere describing any possible pseudospin 
configuration. Thus, the evolution of the out of plane pseudospin polarization $\sigma_z$ is given by the standard relation $\sigma_z(t)=\langle\Phi(0)|U_F^\dagger(t)\sigma_zU_F(t)|\Phi(0)\rangle$,
with $U_F(t)$ being the unitary Floquet evolution operator $U_F(t)=\mathcal{P}^\eta(t)e^{-i\mathcal{H}_Ft}$ (note that 
$\sigma_z$ and $\mathcal{P}^\eta(t)$ commute with each other). 
The initial polarization in the state (\ref{phi0}) is 
given by $\sigma_z(0)=\cos\theta$. After some algebra we find
\begin{eqnarray}\label{zetat}
\nonumber \sigma_z(t)&=&\frac{2\alpha}{\Gamma}\sin\theta\sin\Gamma t\Bigg(\frac{\Delta^\eta_s}{\Gamma}\sin\Gamma t\cos\phi-\cos\Gamma t\sin\phi\Bigg)+\\
&&\cos\theta\Big(1-\frac{2\alpha^2}{\Gamma^2}\sin^2\Gamma t\Big).
\end{eqnarray}
Using this expression, the one-period mean value pseudospin polarization 
\begin{equation}
\langle\sigma_z\rangle=\frac{1}{T}\int^T_0\sigma_z(t)dt, 
\end{equation}
 with $T=2\pi/\Omega$ being the period of oscillations of the driving field, is found to be given as 
\begin{eqnarray}\label{zetapro}
\nonumber\langle\sigma_z\rangle&=&\alpha\sin\theta\Bigg[\frac{\Delta^\eta_ s}{\Gamma^2}\cos\phi\big(1-\sinc(2\Gamma T)\big)-T\sin\phi~\sinc^2(\Gamma T)\Bigg]\\
\nonumber&&+\cos\theta\Bigg[1-\frac{\alpha^2}{\Gamma^2}\Bigg(1-\sinc(2\Gamma T)\Bigg)\Bigg],\\
\end{eqnarray}
where $\sinc(x)=\frac{\sin(x)}{x}$.

\noindent In particular, for  initial states that have zero polarization ($\theta=\pi/2$), we get the simplified expressions
\begin{eqnarray}\label{zetapro0}
\nonumber\langle\sigma_z\rangle&=&\alpha\Bigg[\frac{\Delta^\eta_s}{\Gamma^2}\cos\phi\big(1-\sinc(2\Gamma T)\big)-T\sin\phi\sinc^2(\Gamma T)\Bigg].\\
\end{eqnarray}
Setting the symmetric value $\phi=\pi/4$ and a frequency in the far infrared region $\Omega = 3\textrm{THz}$, we plot in FIG.\ref{fig:figure1} the mean pseudospin polarization for the different spin and valley $s\eta$ product combinations.\\
\noindent From this figure we find that within the low coupling regime ($\alpha\le0.1\Omega$), it is in 
general not possible to induce appreciable changes of the pseudospin polarization and this is related to the fact that the quasienergy behaviour is essentially 
controlled by the parameters $V_z$ and $\lambda_{so}$ which determine the gap behaviour in the static regime. On the other hand, 
for intermediate ($\alpha=0.5\Omega$) and large ($\alpha=0.75\Omega$) values of the coupling to the driving field, i.e., beyond the off resonant condition, effective pseudospin
inversion is achievable and therefore, a qualitatively different behaviour emerges within this coupling regime.\\ 
\noindent The exact results for the pseudospin polarization shown in FIG.\ref{fig:figure1} at vanishing momentum motivate the need to go beyond the off resonant condition for finite values of the particle's momentum,
as we discuss in the following two sections.
\section{Finite momentum: Off resonant regime and beyond} 
\subsection{Quasienergy spectrum}
The dynamics of our system at finite momentum does in general not allow for
a closed analytic solution, and one needs to resort to numerics. A
practical route here is to employ the Fourier expansion of the periodic
part of the Floquet states which turns, after an approprite truncation, the 
Schr\"odinger equation into a finite matrix eigenvalue problem.
Yet, 
before we perform any explicit calculation we physically motivate the need to fully diagonalize the Floquet Hamiltonian, going 
beyond the so called off resonant regime which corresponds to very large frequencies (large compared 
to any other energy scales in the problem) and small coupling strength  presented in reference.\cite{ezawa2} Within this
scheme, the frequency of the driving field is much larger than the unperturbed energy separations. Therefore, only virtual 
single emission-absortion photon processes are allowed. These virtual photons would dress the static eigenstates but could not directly excite 
electronic transitions as happens when real photons are exchanged among the charge carriers. Thus the off resonant and the 
resonant scenarios are clearly physically distinguishable from each other.\\

\noindent For ease of notation let us set $\mathcal{H}_0=\mathcal{H}^\eta$ and $V(t)$ for the static and time-dependent contributions to the full Hamiltonian (\ref{e11t}) which is 
now written as
\begin{equation}\label{general}
\mathcal{H}(t)=\mathcal{H}_0+V(t). 
\end{equation}
Then, within the off resonant approximation we have $\alpha/\Omega\ll1$, and thus one can derive an effective gapped Floquet Hamiltonian (see appendix B. for a detailed derivation)  
\begin{equation}\label{off}
\tilde{\mathcal{H}}_{F}=\mathcal{H}_0+\frac{[V_{-1},V_1]}{\Omega},
\end{equation}
where the $Nth$ harmonic contribution is defined as
\begin{equation}
V_{N}=\frac{1}{T}\int^T_0V(t)e^{-iN\Omega t}dt. 
\end{equation}
The second term in equation (\ref{off}) represents virtual photon emission-absortion processes that would dress the static eigenstates.
Doing the explicit calculation one finds that equation (\ref{off}) becomes
\begin{equation}\label{offfin}
 \tilde{\mathcal{H}}_{F}=\mathcal{H}_0-\eta\frac{\alpha^2}{\Omega}\sigma_z,
\end{equation}
and therefore, a photoinduced modulation of the gap would be possible.\\
\begin{figure*}[ht]
\begin{tabular}{cc}
\includegraphics[height=6.25cm]{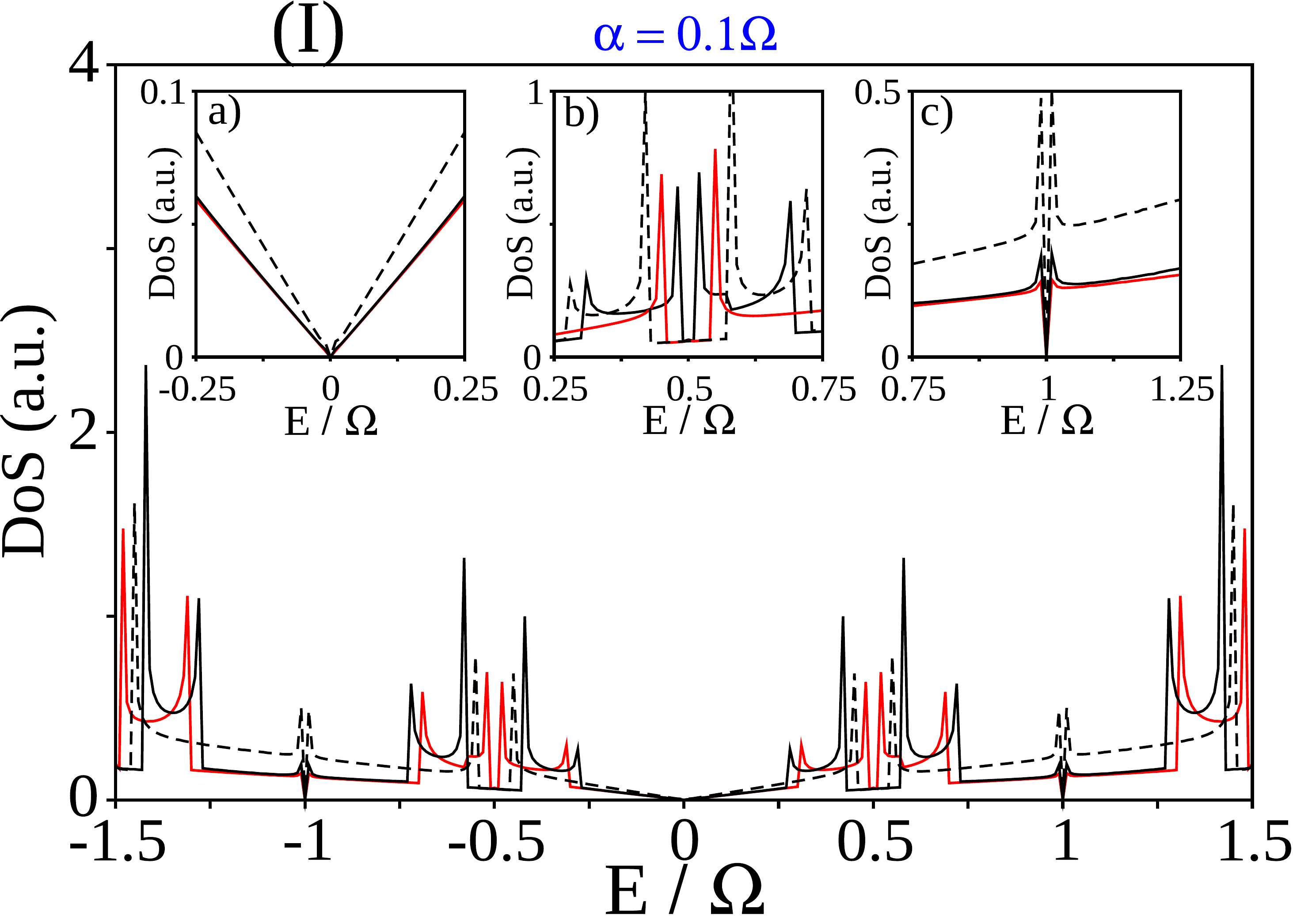}&
\includegraphics[height=6.25cm]{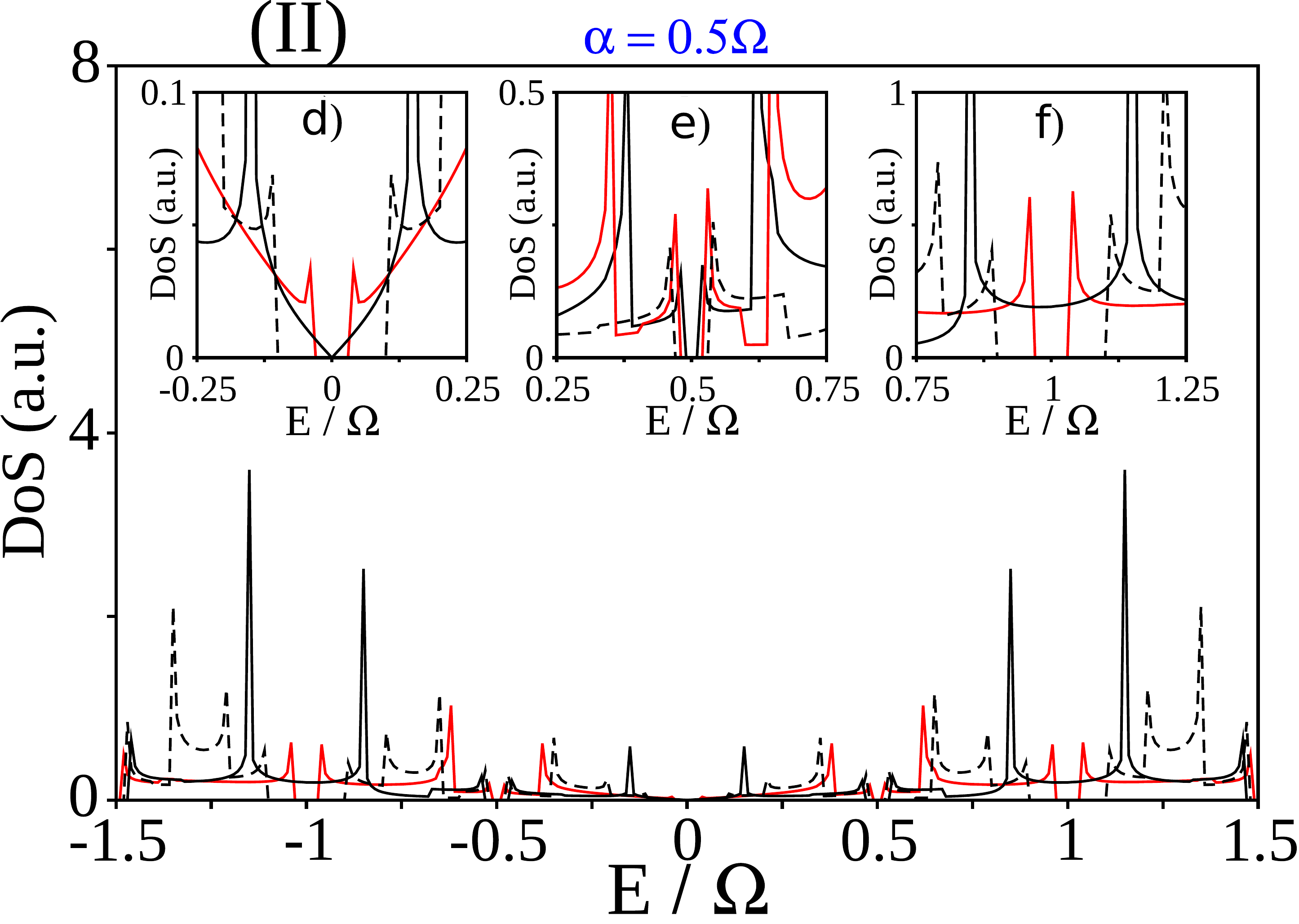}
\end{tabular} 
\caption{\label{fig:figure3} (Color online) Effective coupling dependence of the time-averaged DoS within the low (intermediate) 
coupling regime $\alpha=0.1\Omega$ ($\alpha=0.5\Omega$). Taking as a reference the driven scenario for $\Delta=0$ (black, dashed curve) 
we have set $\Delta=0.4\Omega$ ($\Delta=-0.4\Omega$) for the solid black (red) curve.
The inset (d) of the right panel shows that at intermediate coupling regime one configuration is non gapped (black, continous curve) whereas the other (red curve) is gapped and thus one can achieve the 
driven single-Dirac cone configuration.}
\end{figure*} 
\noindent Yet, under intense terahertz radiation, the conditions that lead to the 
derivation of the last term in eq. (\ref{offfin}) are not satisfied and therefore, appreciable  photoinduced effective gap modulation requires a full tratment of the dynamical 
equations. For instance, if we consider values of the electric field intensities\cite{wu} $\mathcal{E}\sim 0.15\, \textrm{MV/m}$ and  frequencies in the far infrared domain\cite{ganichev} for which 
$\Omega\approx 10\, \textrm{meV}$, one gets for the coupling constant  $\alpha\approx \Omega\approx 10\, \textrm{meV}$ (for a chosen frequency value of $\Omega=3$ THz).  
Therefore, higher order harmonics do contribute and the dynamics must be given a full numerical treatment by Fourier transforming the Schr\"odinger equation, and solving the resulting infinite dimensional static 
eigenvalue problem.\\

\noindent However, we still can get a time-independent 
formulation since the static Hamiltonian (\ref{e1}) commutes with the total angular momentum operator
\begin{equation}
J^\eta_z=xk_y-yk_x+\eta\left(\frac{\sigma_z+s_z}{2}\right). 
\end{equation}
Thus, applying the unitary transformation
\begin{equation}
\mathcal{P}^\eta_z(t)=e^{-iJ^\eta_z\Omega t} 
\end{equation}
we get the effective time-independent Floquet Hamiltonian
\begin{eqnarray}\label{flot}
\mathcal{H}_F&=&\mathcal{H}_0-\Omega J^\eta_z+\alpha\sigma_x.
\end{eqnarray}
This form of the Floquet Hamiltonian is appropriate to evaluate approximate analytical solutions to the dynamics, but we will not follow this semi analytic approach.
Instead, in the following we present numerical solutions to the finite momentum 
dynamics for the coupling regime $\alpha\le\Omega\ll t_b$, with $t_b\approx1.6\, \textrm{eV}$ the hopping parameter in the 
tight binding formulation. \\
 
\noindent Now we present the quasienergy spectra at finite momentum which are obtained by a numerical diagonalization of the periodic Hamiltonian given in equation (\ref{e11t}). 
In FIG.\ref{fig:figure2} we present the 
momentum dependence of the quasienergy spectrum within the low  ($\alpha=0.1\Omega$) and intermediate ($\alpha=0.5\Omega$) 
coupling regimes to the radiation field. Here we have neglected the extrinsic 
spin-orbit contributions $h_{11}$ and $h_{22}$. Yet, the effects of the most important contribution $h_{11}$ can be readily incorporated as it is described in the appendix. We have checked that our 
results do not qualitatively differ when this extrinsic spin-orbit contribution is included. 

\noindent Since the static band structure properties are determined by the sign of the static bandgap $\Delta=\lambda_{so}-V_z$, we have selected two sets of 
significant values of this parameter as it is shown by the red (black), thick curves in FIG.\ref{fig:figure2}  for 
positive (negative) values of the static gap at low $\Delta=0.1\Omega$ and intermediate $\Delta=0.4\Omega$ absolute values of the static gap, respectively. The changes in the static bandgap are controlled through the static electric field 
$E_z$. 
Since the circularly polarized radiation introduces an 
isotropic modulation of the quasienergy spectrum, we can set the value of one of the momentum components, say $k_y=0$ without loss of generality.\\ 

\noindent From the zero energy solution discussed above
we have to take into account that the radiation field also modulates the gap, both through its frequency and amplitude. Therefore,
in order to have a reference for indicating qualitative changes in the band structure we have chosen the quasienergy spectrum for 
$\Delta=0$ (thin lines in FIG.\ref{fig:figure2}). In addition, for finite values of $\Delta$, we use thick arrows that point, for either subband, away from the 
zero bandgap curve signaling how the energy bands are ``pulled away'' in presence of the radiation field. From the results shown in 
FIG.\ref{fig:figure2} we see that at low coupling ($\alpha\le0.1\Omega$), the main modifications of the energy spectrum are 
due to the value of the static bandgap. This is true for both positive (red, thick curves) and negative (black, thick curves) values of
$\Delta$. Yet, at intermediate  values of the light-matter coupling strength ($\alpha=0.5\Omega$), we can infer that the driving field
 is the leading mechanism in modifying the quasienergy spectrum. In fact, as can be seen in the red thick curve 
 (corresponding to $\Delta=0.4\Omega$), at intermediate coupling regime ($\alpha=0.5\Omega$),  the effective bandgap of one of the pseudospin states can be closed at $\Delta=0.4\Omega$. This in turn signals
the onset of the single Dirac cone configuration (red, thick curve in the rightmost panel). However, it is physically distinct in nature to that reported by Ezawa in 
\cite{ezawa2} since it is due to real instead of virtual photon emission and absortion processes. 
\subsection{Density of states} 
To complement the physical picture given before, in this section we present the results for the time-averaged density of states obtained through the expression\cite{andreas}
\begin{equation}
\textrm{DoS}(E)=\sum_{{\bf k},\nu\mu}\sum^\infty_{n=-\infty}\langle\Xi^n_{{\bf k},\mu\nu}|\Xi^n_{{\bf k},\mu\nu}\rangle\delta
[E-\epsilon_{{\bf k},\mu\nu}+n\Omega], 
\end{equation}
where the Floquet eigenstates $|\Xi^n_{{\bf k},\mu,nu}\rangle$ and the quasienergies 
$\epsilon_{{\bf k},\mu\nu}$ are defined via
\begin{equation}
\mathcal{H}_F|\Xi^n_{{\bf k},\mu\nu}\rangle=\epsilon_{{\bf k},\mu\nu}|\Xi^n_{{\bf k},\mu\nu}\rangle. 
\end{equation}

\noindent In FIG.\ref{fig:figure3} we show the resulting time-averaged DoS within the low (intermediate) 
coupling regime $\alpha=0.1\Omega$ ($\alpha=0.5\Omega$) of the Dirac fermions to the radiation field. We have taken as a reference the driven ungapped scenario
$\Delta=0$, shown by the black dashed curve in order to explicitly show the interplay among the driving field and the static 
gap since for $\Delta=0$ no physical configuration of the two pseudospin components  would lead to the single Dirac cone phase.
However, in the inset (d) we can see that for a finite value of $\Delta=0.4\Omega$ and at intermediate
coupling regime, one configuration is non gapped (black, continous curve) for $\Delta=0.4\Omega$, whereas the other 
(red, continous curve) for $\Delta=-0.4\Omega$, is gapped and thus one can achieve the driven single- Dirac cone configuration by properly tuning the ratio of the amplitude/frequency of the driven field at this 
intermediate light-matter coupling values. 

\section{Conclusions}
We have theoretically analyzed the photoinduced effects on a monolayer of silicene subject to intense terahertz circularly polarized electromagnetic radiation.  
We have shown that dynamical gap modulation of the quasienergy spectrum can only occur for large enough coupling strenght regimes of the light-matter interaction 
effective parameter $\alpha$.  We found that for frequencies $\Omega$ within the range of the undriven bandgap real photon emission and absortion
resonant processes induce a ``band inversion`` that changes the qualitative bandgap structure of driven silicene.
Therefore, the intermediate
coupling regime qualitatively reproduces the single Dirac dynamical structure predicted in reference\cite{ezawa2} but with real instead of virtual photon resonant processes and therefore, the obeservation of the physical realization of this topological phase could be achieved at more realistic values of the strength of the light-matter coupling parameter. 
These distinct phases are correlated to the averaged out of plane pseudospin polarization parameter oscillations 
which in turn stem from the angular momentum exchange among the charge carries and the electromagnetic field. 

We would like to add that performing a rotating wave approximation (RWA) would not be suitable to the regime under consideration 
since the corresponding RWA solutions
can only properly describe the dynamics for small values of the coupling constant ($\alpha\approx 0.1\Omega$). We also note that considering another
semianalytical approximation, such as the Magnus expansion\cite{magnus1} could provide some explicit formulae for both the quasienergy spectrum
and Floquet eigenstates. Yet, this approach has the drawback that truncating the series leads to a violation of the stroboscopic
relation which should be a general property of solutions to the dynamics of the periodically driven systems\cite{magnus2}.
From an experimental point of view we consider that the angular momentum exchange between the radiation field and the pseudospin degree of freedom
could be detected by measuring the changes in the polarization state of the reflected radiation from the silicene sample by means of the magneto
optic Kerr effect as it has already been used for detecting real spin effects in semiconducting structures\cite{detection}. 
We consider that our proposed scheme could shed light on the relevance of the pseudospin for practical implementations of this degree of freedom in realistic 
pseudospintronics applications. We would also briefly discuss two additional points that are in order to better understand the physics
of our proposed model. On the one hand, we mention that in order to take into account non-radiative recombination processes 
one should introduce an electron-phonon coupling which was considered in a recent paper by  Mariani and von Oppen\cite{mariani} 
where they have shown that inclusion of this electron-phonon interaction due to transverse or flexural phonos in graphene 
could lead to distinguishable temperature dependences of the single layer graphene resistivity. This is in turn due to the 
fact that flexural phonons dominate the phonon contribution to the resistivity. We could expect that these effects should be 
present in monolayer silicene and would be the focus of future work where one could discuss the interplay between photon and 
phonon couplings to the Dirac fermions in silicene. On the other hand, one could also be interested in  addressing the role 
of scattering effects at finite momenta. In this context, it has been recently shown by Zhai and Jin in reference\cite{zhai} that, within the off 
resonant approximation for epitaxial graphene, the photon dressing of the static eigenstates leads to an assymetry between the 
scattering amplitudes for the inter and intravalley conductances. This is explained as a consequence of the degeneracy 
lifting of the valley degree of freedom which is due to the time reversal symmetry breaking introduced by the electromagnetic 
radiation field. Therefore, we propose that within our setup the pseudospin conductance would have a similar asymmetry but 
the measurability of this asymmetry could be experimentally tested within a more realistic set of parameters since, as we 
have previously discussed in this work, the measurable effects of physical changes within the off resonant assumption 
are far to small to have observable consequences.

\section{Appendix}
\subsection{Block diagonalization of the Hamiltonian}
Following the discussion presented in section III, in this appendix we summarize the block diagonalization procedure of the Hamiltonian to take into account the extrinsic spin-orbit correction $h_{11}$. For simplicity, let us 
 focus on the ${\bf K}$ Dirac point ($\eta=+1$) where we have the $4\times4$ Hamiltonian
\begin{equation}\label{e2}
\mathcal{H}_+(\vec{k})=
\left(
\begin{array}{cccc}
\Delta_-&v_F k_-&iv_2k_-& 0 \\
v_F k_+&-\Delta_-&0& -iv_2k_- \\
-i v_2k_+&0&-\Delta_+& v_F k_-\\
0&iv_2k_+&v_F k_+& \Delta_+ 
\end{array}
\right),
\end{equation}
where $k_\pm=k_x\pm ik_y$, $\Delta_\pm=\lambda_{so}\pm\ell E_z$ and $v_2=a\lambda_{R2}$. 
If we now define $\tan\phi=k_y/k_x$ and perform a unitary transformation with 
\begin{equation}\label{t1}
\tilde{\mathcal{H}}_0(k)=R^\dagger_\phi\mathcal{H}_+(\vec{k})R_\phi 
\end{equation} 
with $R_\phi=\textrm{Diag}(e^{-i\phi},1,1,e^{i\phi})$, we get 
\begin{equation}\label{e3}
\tilde{\mathcal{H}}_0(k)=
\left(
\begin{array}{cccc}
\Delta_-&v_F k&iv_2k& 0 \\
v_F k&-\Delta_-&0& -iv_2k \\
-i v_2k&0&-\Delta_+& v_F k\\
0&iv_2k&v_F k& \Delta_+ 
\end{array}
\right).
\end{equation}
We can further transform the previous Hamiltonian as $\mathcal{H}_0=T^\dagger_\xi\tilde{\mathcal{H}}_0(k)T_\xi$  to get  a block diagonal form 
\begin{equation}\label{1}
\mathcal{H}_0=
 \left(
\begin{array}{cc}
H^-_0(k)& 0 \\
0&H^+_0(k) 
\end{array}
\right),
\end{equation}
where the unitary transformation has the explicit form $T_\xi=\exp(-i\xi\Sigma_0/2)$ and $\xi$ is chosen to get rid of the 
off-diagonal terms. For this purpose we have introduced the $4\times4$ matrix 
\begin{equation}\label{e0}
\Sigma_0=\left(
\begin{array}{cc}
 0& \sigma_0\\
\sigma_0 & 0
\end{array}
\right),
\end{equation}
with $\sigma_0$ the $2\times2$ identity matrix. 
After some straightforward algebra one gets the condition for block diagonalization to fix the angle by the parameter relation 
$\tan\xi=v_2k/\lambda_{so}$. Then, the  diagonal subblocks in Eq.(\ref{1})  read
\begin{equation}\label{emenos}
 H^\pm_0(k)=\mp(\Lambda_k\pm\ell E_z)\sigma_z+v_F k\sigma_x,
\end{equation}
where the effective momentum dependent spin-orbit correction is defined as $\Lambda_k=\sqrt{\lambda^2_{so}+(v_2k)^2}$. 

Under inversion  of the transformation (\ref{t1}), i.e., $R_\phi\mathcal{H}_0R^\dagger_\phi=\mathcal{H}_0(\vec{k})$, we find that 
the upper diagonal subblock of equation (\ref{1})  reads now
\begin{equation}
 H_0(\vec{k})=
 \left(
\begin{array}{cc}
\Delta_k& v_Fke^{-i\phi} \\
v_Fke^{i\phi}&-\Delta_k 
\end{array}
\right),
\end{equation}
where we have simplified the notation by setting the static gap as $\Delta\equiv\Lambda_k-\ell E_z$. 
\subsection{Derivation of the effective Hamiltonian within the off resonant condition}\label{appendixB}
Following the decimation method presented by Medina and Pastawski \cite{medina}, we present now a brief discussion on the derivation of the effective Hamiltonian within the off resonant approximation for the periodic Floquet Hamiltonian $\mathcal{H}(t)=\mathcal{H}_0+V(t)$, as it is given in 
equation (\ref{general}), where $\mathcal{H}_0$ is the static contribution and $V(t+T)=V(t)$ is the time-periodic interaction. Transforming to Fourier space
we get the Floquet Hamiltonian for a monochromatic perturbation in matrix form given in tridiagonal form as
\begin{equation}\left(
 \begin{array}{cccccccccccc}
\ddots\\
\cdots&V_{-1}&\mathcal{H}_{-2}&V_{+1}&0&0&0&\cdots\\
\cdots&0&V_{-1}&\mathcal{H}_{-1}&V_{+1}&0&0&\cdots\\
\cdots&0&0&V_{-1}&\mathcal{H}_0&V_{+1}&0&\cdots\\
\cdots&0&0&0&V_{-1}&\mathcal{H}_{1}&V_{+1}&\cdots\\
&\vdots&\vdots&\vdots&0&V_{-1}&\mathcal{H}_{2}&\cdots\\
&&&&&&&\ddots&&&
\end{array}\right),
\label{floquecino2}
\end{equation}
where the interaction submatrices are defined as
\begin{equation}
 V_{N}=\frac{1}{T}\int^T_0~dtV(t)e^{-iN\Omega t},
\end{equation}
and we have set $\mathcal{H}_N=\mathcal{H}_0+N\Omega$.
If we set out the eigenstate for a given number of Fourier modes $N$ we will have 
\begin{equation}
\Phi=\left(
 \begin{array}{c}
\phi_{-N}\\
\phi_{-N+1}\\
\vdots\\
\phi_{-1}\\
\phi_{0}\\
\phi_{1}\\
\vdots\\
\phi_{N-1}\\
\phi_{N}\\
\end{array}\right),
\end{equation}\\

\noindent with each $\phi_{N}$ being a vector of dimensionality determined by $\mathcal{H}_0$. For instance, if we approximate the problem in such a way that we only consider one Fourier mode ($N=1$), we have to solve
the following system of coupled equations
\begin{eqnarray}
\nonumber&&\mathcal{H}_{-1}\phi_{-1}+V_{+1}\phi_{0}=E\phi_{-1}\\ 
\nonumber&&V_{-1}\phi_{-1}+\mathcal{H}_{0}\phi_{0}+V_{+1}\phi_{+1}=E\phi_{0}\\
\nonumber&&\mathcal{H}_{+1}\phi_{+1}+V_{-1}\phi_{0}=E\phi_{+1}.\\
\end{eqnarray}
From the first and last equations we get
\begin{eqnarray}
\nonumber&&\phi_{-1}=(E-\mathcal{H}_{-1})^{-1}V_{+1}\phi_{0}\\ 
\nonumber&&\phi_{+1}=(E-\mathcal{H}_{+1})^{-1}V_{-1}\phi_{0},\\ 
\end{eqnarray}
such that we get an effective equation for $\phi_0$
\begin{widetext}
\begin{eqnarray}
[V_{-1}(E-\mathcal{H}_{-1})^{-1}V_{+1}+\mathcal{H}_{0}+V_{+1}(E-\mathcal{H}_{+1})^{-1}V_{-1}]\phi_{0}=E\phi_{0}
\end{eqnarray}
\end{widetext}
For $\Omega\gg ||\mathcal{H}_0||$, i.e., frequencies much larger than the typical energy scales of the static problem, we can simplify the denominators
 and approximate 
the previous equation as
\begin{eqnarray}
\left(\mathcal{H}_{0}+\frac{V_{-1}V_{+1}}{\Omega}-\frac{V_{+1}V_{-1}}{\Omega}\right)\phi_{0}\approx E\phi_{0},
\end{eqnarray}
so we get the effective approximate Floquet Hamiltonian, valid for large frequencies
\begin{eqnarray}
\tilde{\mathcal{H}}_\text{F}\approx \mathcal{H}_{0}+\frac{[V_{-1},V_{+1}]}{\Omega}.
\end{eqnarray}
With a similar procedure one can show that for $N=2$ one gets the approximate Floquet Hamiltonian
\begin{eqnarray}
\tilde{\mathcal{H}}^{'}_\text{F}\approx \mathcal{H}_{0}+\frac{[V_{-1},V_{+1}]}{\Omega}-\frac{1}{2\Omega}\frac{[V^2_{-1},V^2_{+1}]}{\Omega^2}.
\end{eqnarray}
{\it{Acknowledgments}--} 
This work has been supported by Deutsche Forschungsgemeinschaft via GRK 1570 and by Yachay EP through a grant from SENESCYT.

\end{document}